\documentstyle[12pt]{article}

\title{\bf  Comment on ``Neutron interferometric observation of
noncyclic phase "}

\vspace{60mm}

\author{\bf Rajendra~Bhandari}

\date{ }

\begin{document}

\maketitle
\vspace{10mm}
\begin{center}
\begin{tabular}{ll}
            & Raman Research Institute, \\
            & Bangalore 560 080, India. \\
            & email: bhandari@rri.ernet.in\\           
\end{tabular}
\end{center}
\vspace{40mm}
PACS NOS.  03.65.Bz , 42.25.Hz, 42.25.Ja

-----------------------------------------------------------------------\\
19 September 1998

\newpage

It was first pointed out in ref.\cite{4pism} that the 
well known neutron interferometer experiments \cite{rauch,werner}
demonstrating the sign change of the wavefunction of odd 
half-integer spin particles under $2\pi$ rotations, done with
unpolarized neutrons, do not
constitute measurement of the phase shift associated with
a given spin state and that such a measurement would require
an experiment with polarized neutrons. It was shown \cite{4pism}
that the continuously monitored phase shift of a spin state
rotating about the polar axis, as given by the Pancharatnam criterion, 
with the initial state taken as the reference state 
(the noncyclic phase), has the
opposite sign for states lying in the upper and the lower
hemispheres and has a discontinuous jump equal to $\pm\pi$
for a state lying on the equator. This was verified experimentally
in optical interference experiments using the polarization 
states of light as a two-state system which is isomorphic
to the spin-1/2 system \cite{4pism,sspsrb}. The phase jumps
in SU(2) evolution occurring at points in the parameter
space where the two interfering states become orthogonal, had
earlier been predicted \cite{jumps,physica} and explained in
terms of jumps in the geometric part of the phase. The origin
of such phase discontinuities in the existence of Dirac
singularities in the parameter space of the SU(2) transformation
was demonstrated,
theoretically and experimentally, for the case of two-state 
system of light polarization \cite{rbdirac,iwbs} and
the occurrence of similar effects in neutron interference
was predicted \cite{4pism,rbdirac,sspsrb,rbreview}.

In ref\cite{wr2}, Wagh et. al. report an experiment
aimed at measuring the noncyclic phase for a spin-1/2
system using neutron interferometry. I wish to point
out that, contrary to the impression
one gets from the paper, the most important part of 
the physics of the noncyclic phase, namely the 
different sign of the phase shift for states in the
upper and the lower hemispheres, is not verified in
this experiment. The reason lies in the fact that
this experiment measures, for whatever
reason,  not the noncyclic phase itself, but a
quantity derived from it, namely the difference of
the phase shift acquired by a given state and the
linear phase shift acquired by the state $\theta = 180^\circ$
or $\theta = 0^\circ$, lying in the other hemisphere.
For example, consider the curves corresponding
to the states $\theta = 70.5^\circ$ and $\theta = 109.5^\circ$
in figure 2 of ref.\cite{wr2}.  For the sake of this argument,
let us ignore the small wiggle in the curves and consider them
as straight lines.
What needs to be shown by measurement is that the 
linear part of the noncyclic phase for states in the upper
hemisphere is $a=\phi/2$ and that for the lower hemisphere
is $b=-\phi/2$, where $\phi$ is the precession angle. 
Instead of measuring a and b, the experiment
shows by measurement that $(a-b)=\phi$ and $(b-a)=-\phi$. The two curves,
therefore, do-not represent independent quantities and none of the 
two implies what needs to be shown. 
It is also noteworthy that the phase shifts plotted in figure 2 
equal $\pm2\pi$ for $2\pi$ rotations on the sphere
and in this sense contain information equivalent to that
in a polarimetric experiment. The quantities $(a-b)$ and $(b-a)$
in fact represent the angles of rotations of the states on
the sphere. In measuring these differences, therefore, one
has sacrificed the true advantage of an
interference experiment in the context of spinor phases.

The curve for $\theta = 90^\circ$ in figure 2 also does
not represent measured quantities. The reason is,  
the points corresponding to $\theta = 90^\circ$ and  
precession angles $\pm 180^\circ$ are phase singularities,
where the phase shift becomes undefined and one is not
justified in making a definite choice for the sign of the
$\pi$ phase jumps as has been done. The choice of sign
in figure 2, which corresponds to choosing $\theta = 90^\circ+\epsilon$,
$\epsilon$ being a small positive quantity, is arbitrary and
does not follow from measured data. 

Another problem with the analysis  of ref.\cite{wr2} is
the use of equation (6) to convert the phase shifts for
partially polarized neutrons to the case of fully polarized
neutrons. An elementary derivation of this equation, not 
reproduced here, shows that this
is valid only for $\theta = 0^\circ$ and $\theta = 180^\circ$
and not in the nonlinear regime.

To conclude, the measurement of the noncyclic phase of an
evolving spinor state alongwith its sign and a demonstration 
of the associated singularity in the case of quantum systems, 
as suggested by the results of the optical polarization experiments,
remains an open question.

\end{document}